\documentclass[sigconf,natbib=false,nonacm,9pt]{acmart}
\microtypesetup{expansion=false}
\AtBeginDocument{%
  \providecommand\BibTeX{{%
    Bib\TeX}}}

\usepackage{amsmath,amssymb,amsfonts}
\usepackage{graphicx}
\usepackage{textcomp}
\usepackage{booktabs}
\usepackage{multirow}
\usepackage{hyperref}
\usepackage{url}
\usepackage{nicefrac}
\usepackage{setspace}
\usepackage{xcolor}
\usepackage{algorithm}
\usepackage{algpseudocode}
\usepackage{tikz}
\usepackage{enumitem}
\usepackage[font=scriptsize,skip=0pt]{caption}
\usetikzlibrary{positioning,arrows.meta,shapes.geometric,calc}

\usepackage{orcidlink}
\usepackage{comment}
\usepackage{braket}
\def\BibTeX{{\rm B\kern-.05em{\sc i\kern-.025em b}\kern-.08em
    T\kern-.1667em\lower.7ex\hbox{E}\kern-.125emX}}

\usepackage{adjustbox}
\usepackage{setspace}
\begin{document}

\title{Investigating Different Barren Plateaus Mitigation Strategies in Variational Quantum Eigensolver}

\author{Mostafa Atallah$^{1,4}$, Nouhaila Innan$^{2,3}$, Muhammad Kashif$^{2,3}$, Muhammad Shafique$^{2,3}$}
\affiliation{$^1$Department of Industrial and Systems Engineering, University of Tennessee - Knoxville, USA\\
$^2$eBRAIN Lab, Division of Engineering, New York University Abu Dhabi (NYUAD), Abu Dhabi, UAE\\
$^3$Center for Quantum and Topological Systems (CQTS), NYUAD Research Institute, Abu Dhabi, UAE\\
$^4$Department of Physics, Faculty of Science, Cairo University, Giza 12613, Egypt
\country{}}

\email{matalla3@vols.utk.edu,nouhaila.innan@nyu.edu,muhammadkashif@nyu.edu,muhammad.shafique@nyu.edu}

\begin{abstract}
Variational Quantum Eigensolver (VQE) algorithms suffer from barren plateaus, where gradients vanish with system size and circuit depth. Although many mitigation strategies exist, their connection to convergence performance under different iteration budgets remains unclear. Moreover, a systematic analysis identifying which state-of-the-art mitigation techniques perform best under specific scenarios is also lacking. We benchmark four approaches, Local-Global, Adiabatic, State Efficient Ansatz (SEA), and Pretrained VQE, against standard VQE on molecular systems from 4 to 14 qubits, analyzing gradient variance up to 50 layers and convergence over 1000 iterations.
Our results show that the impact of gradient preservation is iteration-dependent. In the 14-qubit BeH$_2$ system, Pretrained VQE outperforms SEA at 100 iterations despite lower gradient variance, but SEA becomes 2.2$\times$ more accurate at 1000 iterations. For smaller systems, SEA achieves near-exact energies (H$_2$: $10^{-5}$ Ha, LiH: $2\times10^{-4}$ Ha) with fidelities 0.999, while standard methods plateau early.
The results demonstrate that robust barren plateau mitigation depends on aligning the chosen strategy with both system size and available computational budget, rather than treating gradient variance as the sole predictor of performance.
\end{abstract}

\maketitle
\begin{spacing}{0.97}

\section{Introduction}
Variational Quantum Eigensolver (VQE) represents a promising hybrid quantum-classical
approach for ground state energy estimation in molecular systems. By combining
parameterized quantum circuits with classical optimization, VQE offers a near-term
pathway to quantum advantage in computational chemistry \cite{shang2023towards}.
However, it faces a fundamental challenge in the form of barren plateaus (BPs), where
gradients vanish exponentially with system size and circuit depth, rendering the
optimization infeasible.
BPs emerge from the interplay between quantum circuit expressivity and trainability.
While deep, highly-entangled circuits can represent complex quantum states, they
simultaneously create exponentially flat optimization landscapes
\cite{mcclean2018barren,kashif2024resqnets}. Numerous mitigation strategies have been proposed aiming to overcome the gradient vanishing issue by preseving gradient's variance throughout the optimization process. However, the effectiveness of these approaches across multiple molecular systems, circuit depths, and performance metrics is largely unexplored.
\begin{figure}
    \centering
    \includegraphics[width=1.0\linewidth]{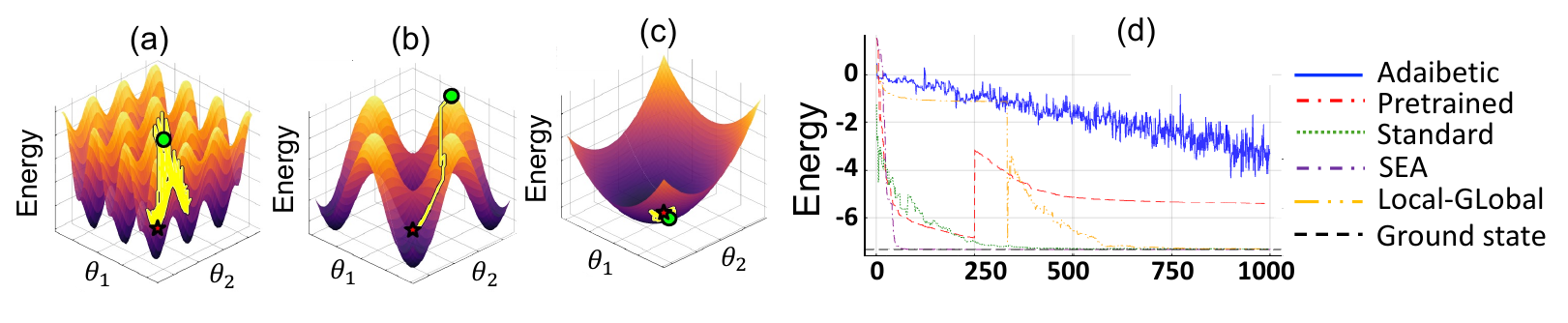}
    \caption{Impact of BP mitigation strategies on optimization landscape (a) SEA, (b) Adiabatic initialization, (c) Pretrained VQE, and (d) VQE convergence across different methods.}
    \label{fig:mot_study_BP}
\end{figure}

\textbf{Motivational Analysis:}
We tested different state-of-the-art solutions to BeH$_2$ molecule with same size of VQE ansatz consisting of $14$ qubits and depth of $30$ layers. We observe that different solutions lead to different  optimization landscape, as shown in Fig. \ref{fig:mot_study_BP}(a,b,c). This highlights the needs to perform a comprehensive analysis of how effective different techniques are, and under what scenarios. Similarly, the convergence behavior in different state-of-the-art solution is different (Fig. \ref{fig:mot_study_BP}(d)), which further highlights the need for comprehensive investigation into how the gradient's variance is related to overall convergence.

Based on our observations, we explore two important questions in this paper: (1) How the different state-of-the-art BPs solutions works across different settings? (2) Does preserving the gradient variance guaranty better convergence?
To this end, we systematically benchmark four BPs mitigation strategies against standard VQE (without any mitigation strategy applied), across different molecular Hamiltonians requiring different numbers of qubits. \textbf{Our novel contributions are summarized below:}
\begin{figure*}[htbp]
    \centering
    \includegraphics[width=1.0\linewidth]{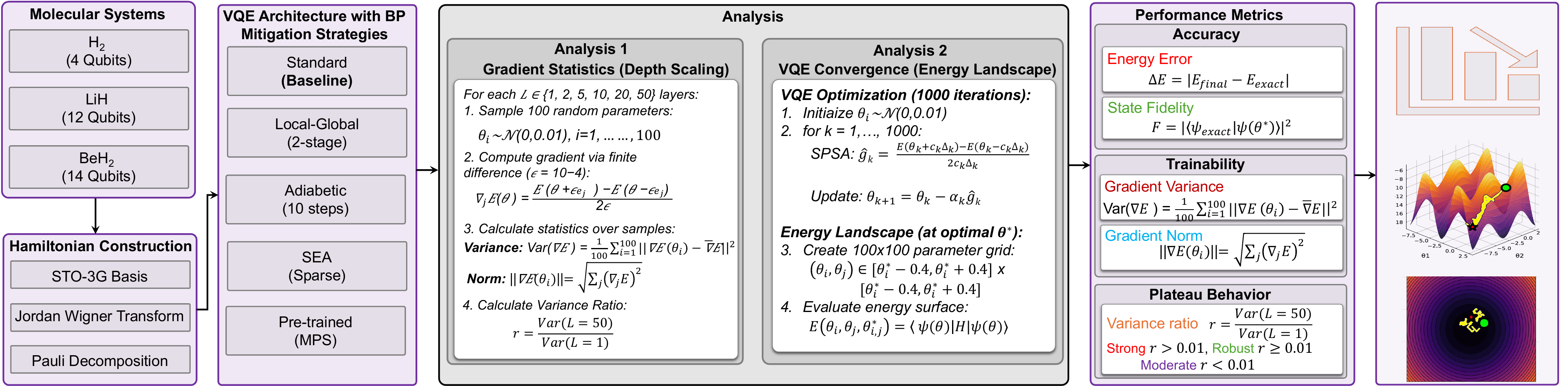}
    \caption{Methodology pipeline for benchmarking barren plateau mitigation strategies in VQE. The workflow progresses from molecular system specification through Hamiltonian construction, VQE optimization with five methods, to comprehensive evaluation using gradient variance analysis, convergence metrics, and loss landscape visualization.}
    \label{fig:methodology_pipeline}
\end{figure*}

\begin{itemize}[leftmargin=*]
    \item A comprehensive empirical comparison of four mitigation strategies across systems ranging from 4 to 14 qubits with circuit depths of up to 50 layers.

    \item Demonstration that the relationship between gradient variance and convergence success is fundamentally iteration-dependent, with initialization quality providing short-term advantage while gradient preservation determines long-term optimization capacity.

    \item Evidence that SEA VQE achieves near-exact convergence for small-to-medium systems when given sufficient iterations despite using a constrained ansatz.

    \item Practical recommendations for selecting mitigation strategies based on system size and computational budget rather than gradient variance alone.
\end{itemize}

\section{Background and Related Work}
\label{background}

\subsection{Variational Quantum Eigensolver}

The VQE algorithm minimizes the energy expectation value of a molecular Hamiltonian $H$ through the expression {\footnotesize $E(\theta) =\langle \psi(\theta) | H | \psi(\theta) \rangle$, where $|\psi(\theta)\rangle = U(\theta)|0\rangle$} is prepared by applying a parameterized ansatz $U(\theta)$ to the initial state. The algorithm proceeds iteratively, with quantum hardware evaluating $E(\theta)$ for given parameters while a classical optimizer updates $\theta$ to minimize energy \cite{peruzzo2014variational,tilly2022variational,chen2024crossing}. This hybrid quantum-classical approach makes VQE particularly suitable for noisy intermediate-scale quantum (NISQ) devices \cite{cerezo2021variational,innan2024quantum}.

The ansatz $U(\theta)$ typically consists of alternating layers of single-qubit rotations and entangling gates. For $L$ layers and $n$ qubits, the parameter count scales as $\mathcal{O}(nL)$, creating high-dimensional optimization landscapes. The choice of ansatz architecture critically impacts both expressivity and trainability, with hardware-efficient designs balancing gate count against representational power \cite{wecker2015progress,boutakka2025benchmarking}. Recent reviews \cite{fedorov2022vqe} provide comprehensive coverage of VQE methods and best practices for practical applications.

\subsection{Barren Plateaus}

BPs occur when the variance of partial derivatives of the cost function vanishes with increasing system size or circuit depth \cite{mcclean2018barren,kashif2024alleviating}. This makes gradient-based optimization infeasible.

The physical origins of BPs stem from several interconnected mechanisms
\cite{larocca2024review,larocca2025barren}, such as, Global cost functions that
measure observables acting on all qubits, causing gradients to concentrate in
exponentially small regions of the parameter space
\cite{cerezo2021cost,kashif2023impact}. Deep circuits with strong entanglement create
effective averaging over exponentially large Hilbert spaces, resulting in a diluted
gradient information \cite{kashif2025deep}.
Generic random initialization leads to exploration of parameter spaces containing exponentially many local minima with similar cost function values \cite{grant2019initialization,kashif2024dilemma}.
Noise in quantum hardware can further exacerbate BPs \cite{wang2021noise,kashif2024hqnet,kashif2024nrqnn}, creating additional challenges for practical implementations. Beyond vanishing gradients, BPs exhibit characteristic signatures including flat loss landscapes, high sensitivity to parameter initialization, and poor scaling behavior with both system size and circuit depth \cite{larocca2022diagnosing,fontana2024characterizing}.

\subsection{BP Mitigation Strategies}
Extensive research has explored various approaches to mitigate BPs. Below we highlight some notable techniques, which are also used in this paper.

\begin{enumerate}[leftmargin=*]
    \item \textbf{Local-Global VQE} \cite{cerezo2021cost} implements a two-stage optimization strategy. In the first stage, the algorithm optimizes on a local Hamiltonian constructed by extracting only single-qubit and nearest-neighbor interaction terms from the full molecular Hamiltonian. The optimized parameters from this local stage then serve as initialization for the second stage, which performs optimization on the complete global Hamiltonian. The rationale is that local Hamiltonians exhibit reduced BPs effects, providing a more favorable starting point for global optimization.

    \item \textbf{Adiabatic VQE} \cite{li2017efficient} gradually transitions between Hamiltonians through interpolation. The method constructs a sequence of interpolated Hamiltonians $H(s) = (1-s)H_{\text{local}} + sH_{\text{global}}$ where $s=[0,1]$. The optimization is divided into multiple adiabatic steps, with each step optimizing the Hamiltonian at a particular interpolation value while warm-starting from the previous optimum. This approach aims to provide smooth parameter space evolution that avoids abrupt landscape changes characteristic of direct global optimization.

    \item \textbf{State Efficient Ansatz VQE} \cite{liu2205mitigating} employs a carefully structured ansatz with constrained depth and sparse entanglement patterns. The implementation uses a depth configuration controlling rotation layers in three sections, with reduced connectivity between qubits compared to standard hardware-efficient anstaz. By limiting circuit expressivity, this approach seeks to maintain gradient trainability while preserving sufficient representational power for accurate ground state approximation.

    \item \textbf{Pretrained VQE} \cite{dborin2022matrix} uses a two-stage approach with Matrix Product State initialization. The first stage trains an MPS-inspired ansatz featuring sequential nearest-neighbor entanglement structure. Parameters from this MPS optimization are then transferred to initialize a more expressive full ansatz, which undergoes further optimization in the second stage. The MPS structure provides physically-motivated initialization that may position parameters in favorable regions of the optimization landscape.
\end{enumerate}

\section{Methodology}

We comprehensively evaluate state-of-the-art BP mitigation strategies. Our pipeline spans the full VQE workflow, from molecular problem formulation to multi-metric performance analysis. We construct three molecular Hamiltonians for H$_2$, LiH, and BeH$_2$ using the STO-3G basis, Jordan--Wigner mapping, and Pauli decomposition. We then incorporate four different BP mitigation strategies and perform a comparative analysis of these technqiues with standard VQE.
To quantify the effectiveness of these approaches, we perform two analyses: (i) gradient-scaling analysis, where we measure depth-dependent gradient propagation statistics to analyze susceptibility to BPs, and (ii) convergence and landscape analysis, where we study optimization landscapes and the local energy topology around the converged solution using SPSA-based VQE runs. Finally, we evaluate each method w.r.t accuracy, trainability, and plateau-strength, including energy error, fidelity, gradient variance, gradient norm, and variance ratios across depths. Fig.~\ref{fig:methodology_pipeline} illustrates the complete workflow.

\vspace{-7pt}
\subsection{Molecular Systems}
We selected three molecules to probe barren plateau transition regimes: \textbf{H$_2$} (0.735 \AA{} bond length, 2 electrons, 4 qubits) as minimal baseline with no BP effects; \textbf{LiH} (1.595 \AA{}, 4 electrons, 12 qubits) exhibiting moderate BP behavior; and \textbf{BeH$_2$} (1.33 \AA{} Be-H bonds, 6 electrons, 14 qubits) with strong BP characteristics. This progression systematically reveals how mitigation effectiveness depends on system size.
\vspace{-7pt}

\subsection{Hamiltonian Construction}
We employ STO-3G minimal basis yielding 2, 6, and 7 spatial orbitals for H$_2$, LiH, and BeH$_2$ respectively, with full configuration interaction (no active space truncation). The electronic Hamiltonian
\begin{equation}
H = \sum_{pq} h_{pq} a_p^\dagger a_q + \frac{1}{2}\sum_{pqrs} h_{pqrs} a_p^\dagger a_q^\dagger a_r a_s + E_{\text{nuc}},
\end{equation}
is mapped to qubits via Jordan-Wigner transformation

$a_j^\dagger \rightarrow \frac{1}{2}(\prod_{k<j} Z_k)(X_j - iY_j)$, producing $H_{\text{qubit}} = \sum_i c_i P_i$, where $c_i$ are real coefficeints and $P_i$ is a Pauli operator.
Exact FCI ground states provide reference energies: H$_2$ ($-7.309$ Ha), LiH ($-19.110$ Ha), BeH$_2$ ($-36.665$ Ha).

\subsection{VQE Architecture with BP Mitgation}

\textbf{For Standard, Local-Global, and Adiabatic VQE}, the ansatz design uses the EfficientSU2 structure consisting of alternating rotation and entangling layers. Each rotation layer applies Ry and Rz gates to all qubits, while entangling layers use circular CNOT connectivity.
For $L$ layers, the circuit applies the repeated block {\footnotesize
$U(\theta) = \prod_{l=1}^L \left[ R_{\text{ent}} \prod_{i=1}^n R_y(\theta_{i,l}^y) R_z(\theta_{i,l}^z) \right]$,} where $R_{\text{ent}}$ applies circular CNOT gates connecting qubit $i$ to qubit $(i+1) \bmod n$. The total number of parameters equals $2nL + 2n$, accounting for two rotation gates per qubit per layer plus a final rotation layer.

\textbf{SEA VQE} employs a simplified ansatz structure with reduced expressivity designed specifically for BPs mitigation. The ansatz uses a depth configuration $[d_1, d_2, d_3] = [1, 1, 1]$ controlling rotation layers in three distinct sections:
{\scriptsize $U_{\text{SEA}}(\theta) = R_{\text{full}} \prod_{d=1}^{d_3}$  $\left[\prod_{i=1}^n R_y(\theta_{i,d}^{(3)}) \right] R_{\text{cross}} \prod_{d=1}^{d_2} \left[\prod_{i=1}^n R_z(\theta_{i,d}^{(2)}) \right]
R_{\text{sparse}} \prod_{d=1}^{d_1}$  $\left[\prod_{i=1}^n R_y(\theta_{i,d}^{(1)}) \right], \quad\quad$} where $R_{\text{sparse}}$ applies CNOTs connecting only odd-indexed qubits to their immediate neighbors (qubits 1$\rightarrow$2, 3$\rightarrow$4, etc.), creating partial entanglement; $R_{\text{cross}}$ applies CNOTs linking qubits separated by two positions (qubits 1$\rightarrow$3, 3$\rightarrow$5, etc.), enabling medium-range correlations; and $R_{\text{full}}$ connects the first half of qubits to the second half (qubit $i$ to qubit $i + \lfloor n/2 \rfloor$ for $i \leq \lfloor n/2 \rfloor$), establishing global correlations while avoiding exponential entanglement growth. This sparse connectivity pattern limits entanglement growth while maintaining sufficient expressivity for smaller molecules. For $n$ qubits and depth configuration $[d_1, d_2, d_3]$, the parameter count equals $n(d_1 + d_2 + d_3)$. At 50 layers, SEA VQE uses 2100 parameters for BeH$_2$ (14 qubits) compared to 1428 parameters for standard methods.

\textbf{Pretrained VQE} uses a two-stage architecture with distinct Ans\"atze for each phase. The first stage employs an MPS-inspired ansatz with bond dimension 2, featuring sequential nearest-neighbor entanglement structure:
{\footnotesize $U_{\text{MPS}}(\theta) = \prod_{i=1}^{n-1} \left[ \text{CNOT}_{i,i+1} \, R_y(\theta_i) \, \text{CNOT}_{i,i+1} \right].$}
The MPS ansatz begins with single-qubit Ry and Rz rotations on all qubits, followed by forward and backward sweeps of two-qubit operations. For bond dimension $\chi=2$, the MPS stage uses $2n + 2(n-1)\chi$ parameters. The second stage transitions to the full EfficientSU2 ansatz with $2nL + 2n$ parameters. Parameter transfer maps the optimized MPS parameters to the first $\min(n_{\text{MPS}}, n_{\text{full}})$ positions of the full ansatz parameter vector, with remaining parameters initialized from a small-scale normal distribution with standard deviation 0.01. This transfer preserves the physical structure learned during MPS optimization while allowing the full ansatz to refine the solution.
\vspace{-10pt}
\subsection{Analysis 1: Gradient Statistics}

To quantify how barren plateau severity evolves with circuit depth, we systematically probe the gradient landscape through random sampling across six depth configurations: $L \in \{1, 2, 5, 10, 20, 50\}$ layers. This analysis captures the transition from shallow, trainable circuits to deep architectures where gradient information may vanish exponentially.

For each depth $L$ and VQE method, we draw $N_{\text{samples}} = 100$ random parameter configurations $\{\theta^{(i)}\}_{i=1}^{N_{\text{samples}}}$ uniformly from $(0, 2\pi)$. These random samples explore the parameter space independent of any optimization trajectory, providing an unbiased view of typical gradient magnitudes encountered during training.

At each sampled configuration $\theta^{(i)}$, we estimate the gradient components via central finite differences. For parameter index $j$, we perturb the configuration by a small step $\epsilon = 10^{-4}$ along the $j$-th direction and compute
\begin{equation}
[\nabla_\theta E(\theta^{(i)})]_j = \frac{E(\theta^{(i)} + \epsilon e_j) - E(\theta^{(i)} - \epsilon e_j)}{2\epsilon},
\end{equation}
where $e_j$ is the $j$-th unit vector and $E(\theta) = \langle \psi(\theta) | H | \psi(\theta) \rangle$ is the quantum expectation value. This symmetric difference scheme provides second-order accuracy while requiring only two energy evaluations per parameter. Aggregating these components, we obtain the gradient norm $\|\nabla E(\theta^{(i)})\| = \sqrt{\sum_{j=1}^{d} [\nabla_\theta E(\theta^{(i)})]_j^2}$, which quantifies the magnitude of the energy landscape's slope at point $\theta^{(i)}$.

The distribution of these gradient norms across all samples reveals the trainability landscape. We characterize this distribution through its variance
\begin{equation}
\text{Var}(\nabla E, L) = \frac{1}{N_{\text{samples}}} \sum_{i=1}^{N_{\text{samples}}} \left(\|\nabla E(\theta^{(i)})\| - \overline{\nabla E}\right)^2,
\end{equation}
where $\overline{\nabla E} = \frac{1}{N_{\text{samples}}} \sum_{i=1}^{N_{\text{samples}}} \|\nabla E(\theta^{(i)})\|$ represents the mean gradient norm. High variance indicates that gradients vary substantially across the parameter space, providing diverse directional information that guides optimization. Conversely, low variance approaching zero signals that gradients have become uniformly small, indicating the presence of BPs.
\vspace{-7pt}
\subsection{Analysis 2: VQE Convergence}

While gradient variance reveals trainability in principle, actual ground state approximation quality depends on the full optimization dynamics. To bridge this gap, we execute complete VQE runs for each method and molecule, tracking how trainability translates into convergence behavior and final accuracy. We fix the circuit depth at $L = 30$ layers---sufficiently deep to exhibit barren plateau effects while remaining expressive enough to approximate ground states accurately. The iteration budget is set to $N_{\text{max}} = 1000$, and parameters initialize from a narrow Gaussian $\theta_0 \sim \mathcal{N}(0, 0.01)$ with fixed random seed 42 to ensure reproducibility across all experiments.

The optimization engine employs Simultaneous Perturbation Stochastic Approximation (SPSA), a gradient-free method particularly suited for noisy optimization landscapes. SPSA adapts both the learning rate and perturbation magnitude over iterations through power-law schedules: $\alpha_k = a/(k+1)^{\alpha}$ controls the step size with $a=0.1$ and $\alpha=0.602$, while $c_k = c/(k+1)^{\gamma}$ governs perturbation decay with $c=0.1$ and $\gamma=0.101$. These coefficients follow standard calibration ensuring convergence guarantees under mild regularity conditions.

At each iteration $k$, SPSA samples a random perturbation direction $\Delta_k$ with components drawn uniformly from $\{-1, +1\}$, then evaluates the energy at two symmetric points $\theta_k \pm c_k \Delta_k$. The gradient estimate follows from the finite difference
\begin{equation}
\hat{g}_k = \frac{E(\theta_k + c_k \Delta_k) - E(\theta_k - c_k \Delta_k)}{2c_k \Delta_k},
\end{equation}
where $E(\theta) = \langle \psi(\theta) | H | \psi(\theta) \rangle$. This simultaneous perturbation of all parameters yields computational efficiency: estimating the full gradient requires only two function evaluations regardless of parameter count, compared to $2d$ evaluations for finite differences. The parameter update then proceeds via gradient descent $\theta_{k+1} = \theta_k - \alpha_k \hat{g}_k$, steering the parameters toward lower energy configurations.

\subsection{Performance Metrics and Evaluation Criteria}

We evaluate mitigation strategies through three metric categories.
\textit{Accuracy metrics} quantify ground state approximation quality: energy error
$\Delta E = |E_{\text{final}} - E_{\text{exact}}|$ measures deviation from exact FCI
energies (threshold: 0.1 Ha), while state fidelity
$F = |\langle \psi_{\text{exact}} | \psi(\theta^*) \rangle|^2$ captures wavefunction
overlap ($F < 0.99$ reveals discrepancies energy alone might miss).
\textit{Trainability metrics} assess optimization feasibility: gradient
norm $\|\nabla E(\theta_i)\| = \sqrt{\sum_{j} [\nabla_j E]^2}$ quantifies landscape
slope magnitude, and gradient variance
$\text{Var}(\nabla E) = \frac{1}{N}\sum_{i=1}^{N} (\|\nabla E(\theta_i)\| - \overline{\nabla E})^2$ measures directional information diversity across parameter space.
\textit{Plateau behavior metrics} classify mitigation effectiveness: variance ratio
$R = \text{Var}(\nabla E, L=50) / \text{Var}(\nabla E, L=1)$ partitions methods into
regimes (\textit{Gradient Maintained} $R \geq 0.5$; \textit{Moderate Plateau}
$0.01 \leq R < 0.5$; \textit{Strong Plateau} $R < 0.01$).

\vspace{-5pt}
\section{Results and Discussion}

\subsection{Experimental Setup}

All experiments use a maximum iteration count of 1000 for VQE optimization.
Initial parameters are drawn from a normal distribution with mean zero and standard deviation 0.01,
using a fixed random seed of 42 for reproducibility. The SPSA optimizer parameters are set
to $a = 0.1$, $c = 0.1$, $A = 0$, $\alpha = 0.602$, and $\gamma = 0.101$.
Layer depths tested include 1, 2, 5, 10, 20, and 50 layers for the scaling analysis. For the detailed
30-layer analysis, 1000 iterations are used with landscape resolution of 100$\times$100 grid points over a
range of $\pm 0.4$ radians. For gradient variance analysis, 100 parameter samples are evaluated per depth
configuration with finite difference epsilon equal to $10^{-4}$. Energy evaluations are performed through
exact state vector simulation without shot noise.
\vspace{-7pt}
\subsection{Gradient Variance Analysis}
To evaluate trainability, we examine how gradient variance changes with circuit depth for all methods and molecules.
\begin{table}[htpbt]
    \small
\centering
\caption{Gradient variance scaling across molecular systems}
\label{tab:variance_scaling}
\setlength{\tabcolsep}{3pt}
\renewcommand{\arraystretch}{1.1}
\begin{adjustbox}{max width=1.0\linewidth}
\begin{tabular}{l l c c c l}
\toprule
Molecule & Method & Var$_\text{1L}$ & Var$_\text{50L}$ & Var. Ratio & Classification \\
\midrule

\multirow{5}{*}{H$_2$}
& Standard. & 0.085 & 0.147 & 1.726 & Grad. Maintained \\
& Local-Global & 0.066 & 0.156 & 2.367 & Grad. Maintained \\
& Adiabetic & 0.074 & 0.149 & 2.008 & Grad. Maintained \\
& Pretrained & 0.071 & 0.141 & 2.002 & Grad. Maintained \\
& SEA & 0.370 & 0.219 & 0.592 & Grad. Maintained \\
\midrule

\multirow{5}{*}{LiH}
& Std. & 0.088 & 0.00209 & 0.024 & Mod. Plateau \\
& Local-Global & 0.073 & 0.00206 & 0.028 & Mod. Plateau \\
& Adiabetic & 0.081 & 0.00208 & 0.026 & Mod. Plateau \\
& Pretrained & 0.087 & 0.00207 & 0.024 & Mod. Plateau \\
& SEA & 0.459 & 0.318 & 0.692 & Grad. Maintained \\
\midrule

\multirow{5}{*}{BeH$_2$}
& Std. & 0.236 & 0.00165 & 0.0070 & Strong Plateau \\
& Local-Global & 0.221 & 0.00165 & 0.0074 & Strong Plateau \\
& Adiabetic & 0.238 & 0.00166 & 0.0070 & Strong Plateau \\
& Pretrained & 0.216 & 0.00164 & 0.0076 & Strong Plateau \\
& SEA & 1.291 & 0.788 & 0.611 & Grad. Maintained \\
\bottomrule
\end{tabular}
\end{adjustbox}
\end{table}
As shown in Table~\ref{tab:variance_scaling}, for H$_2$, all methods maintain high gradient variance at 50 layers, with several showing ratios greater than one. This indicates that barren plateau effects do not emerge in this small system, and gradient norms remain stable or increase with depth.

For LiH, moderate barren plateau behavior is observed. Standard, Local-Global, Adiabatic, and Pretrained VQE exhibit sharp variance decay (ratios 0.024--0.028). SEA VQE is the only method that maintains substantial variance (ratio 0.692), with gradient norms continuing to increase while the others flatten.

For BeH$_2$, strong barren plateau behavior appears for all standard methods, with variance collapsing by more than two orders of magnitude (ratios $\sim$0.007). SEA VQE again preserves significantly higher variance (ratio 0.611), resulting in gradient magnitudes over 25$\times$ larger than the other methods at deep circuit depths.

\vspace{-7pt}
\subsection{Convergence Energy Analysis}
\begin{figure*}[htpb]
    \centering
    \includegraphics[width=1\linewidth]{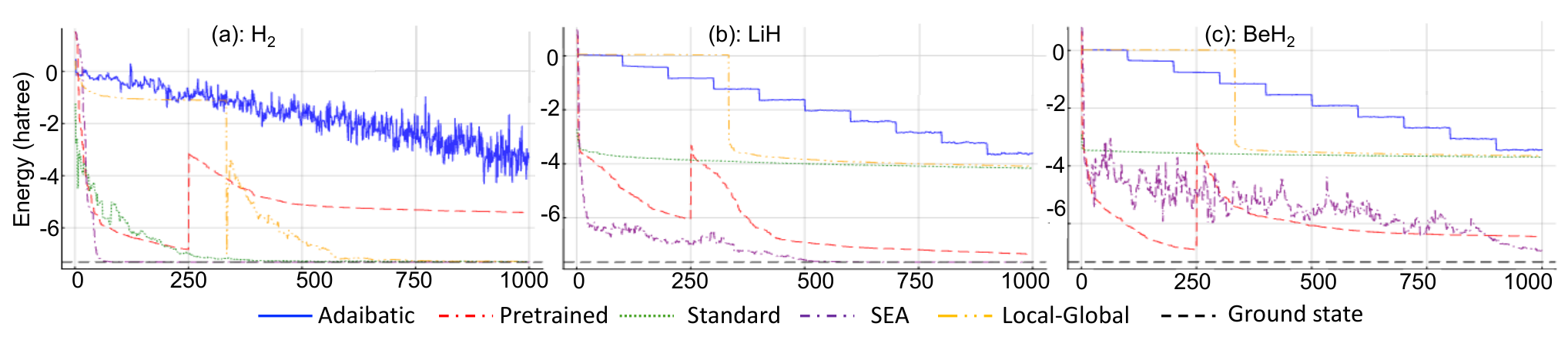}
    \vspace{-0.4cm}
    \caption{Energy convergence for all five methods on H$_2$ (4 qubits), (b) LiH (12 qubits), and (c) BeH$_2$ (14 qubits). All experiments use 1000 iterations at 30 layers depth.}
    \label{fig:convergence}
\end{figure*}
For H$_2$ (exact energy $-7.309$ Ha), as shown in Fig.~\ref{fig:convergence}-a, SEA VQE reaches near-exact accuracy with final energy $-7.309$ Ha (error $1\times10^{-5}$ Ha, fidelity 0.9999). Standard and Local-Global VQE converge close to the ground state with errors of 0.018 Ha and 0.017 Ha, respectively. Pretrained VQE yields an error of 1.892 Ha, while Adiabatic VQE performs poorly with a 4.558 Ha error. Execution time varies considerably: Pretrained (0.687 s) and SEA (0.825 s) are much faster than Standard VQE (23.445 s).

For LiH (exact energy $-19.110$ Ha), as shown in Fig.~\ref{fig:convergence}-b, SEA VQE again achieves near-exact convergence (error 0.0002 Ha, fidelity 0.9999). Pretrained VQE improves substantially compared to short-horizon optimization, reaching $-18.378$ Ha (0.733 Ha error). The remaining methods plateau far from the ground state despite the extended iteration budget: Local-Global yields an 8.883 Ha error, Standard VQE 8.692 Ha, and Adiabatic VQE 10.115 Ha. Runtime ranges from 6.075 s for SEA to 203.473 s for Standard VQE.

BeH$_2$ (exact energy $-36.665$ Ha) reveals the most notable trends, as illustrated in Fig.~\ref{fig:convergence}-c. SEA VQE obtains the best final accuracy at $-34.692$ Ha (1.973 Ha error), a 7.1$\times$ improvement over its 100-iteration result. Pretrained VQE achieves $-32.242$ Ha (4.423 Ha error), improving 2.2$\times$ over its shorter run. Standard, Local-Global, and Adiabatic VQE remain trapped in high-energy regions with errors above 18 Ha, showing little improvement after the first 100--200 iterations.

A key outcome is the ranking reversal between SEA and Pretrained VQE for BeH$_2$. At 100 iterations, Pretrained VQE outperforms SEA (errors 9.648 Ha vs.\ 13.967 Ha), but after 1000 iterations SEA achieves a significantly lower error (1.973 Ha vs.\ 4.423 Ha). This demonstrates that initialization offers short-term benefit, whereas preserved gradient variance governs long-term optimization capability. The three standard VQE variants show negligible improvement beyond early iterations, indicating that vanishing gradients fundamentally limit their performance regardless of iteration budget.

\vspace{-10pt}
\subsection{Loss Landscapes}
To characterize the optimization behavior of each VQE method, we examine the loss landscapes at 30 layers and quantify convergence stability via the standard deviation of energy values over the final 20 iterations.
\begin{figure*}
    \centering
    \includegraphics[width=1\linewidth]{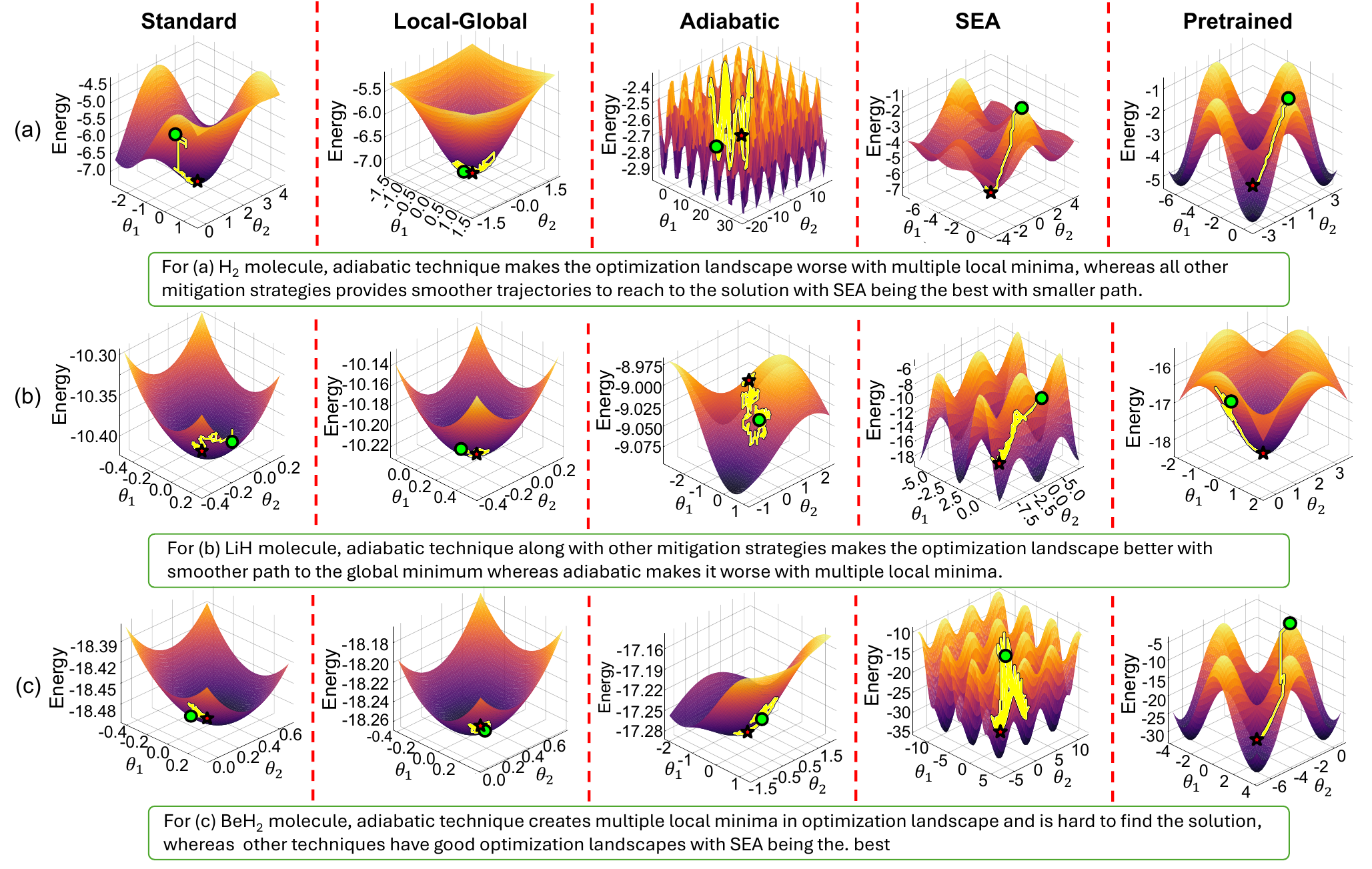}
    \caption{Loss landscapes (30 layers) for (a) H$_2$, (b) LiH (b), and (c) BeH$_2$.}
    \label{landscapes}
\end{figure*}

For H$_2$, the 3D landscapes in Fig.~\ref{landscapes}-a show clear differences among the methods. SEA VQE forms a sharply defined minimum with the lowest stability value ($3.52 \times 10^{-6}$), and its gradient norm (0.590) and fidelity (0.9999) indicate near-exact convergence. Standard and Local-Global VQE also converge well, with stabilities of 0.0014 and 0.0010, respectively. In contrast, Adiabatic VQE exhibits a much flatter basin (stability 0.417), consistent with its poor energy accuracy, while Pretrained VQE shows moderate structure.
\begin{table}[htpb]
    \small
\centering
\caption{Loss Landscape Characteristics at 30 Layers (1000 iterations)}
\label{tab:landscape_stability}
\begin{adjustbox}{max width=1.0\linewidth}
\begin{tabular}{@{}lcccc@{}}
\toprule
Molecule & Method & Stability & Grad. Norm & State Fidelity \\
\midrule
\multirow{5}{*}{H$_2$}
& Standard & 0.0014 & 7.460 & 0.998 \\
& Local-Global & 0.0010 & 6.708 & 0.998 \\
& Adiabatic & 0.417 & 5.982 & 0.536 \\
& Pretrained & 0.0032 & 0.859 & 0.772 \\
& SEA & $3.52 \times 10^{-6}$ & 0.590 & 0.9999 \\
\midrule
\multirow{5}{*}{LiH}
& Standard & 0.0076 & 1.291 & 0.635 \\
& Local-Global & 0.0033 & 1.274 & 0.628 \\
& Adiabatic & 0.0495 & 1.217 & 0.589 \\
& Pretrained & 0.0108 & 1.494 & 0.962 \\
& SEA & $1.22 \times 10^{-5}$ & 1.041 & 0.9999 \\
\midrule
\multirow{5}{*}{BeH$_2$}
& Standard & 0.0076 & 1.217 & 0.609 \\
& Local-Global & 0.0058 & 1.218 & 0.605 \\
& Adiabatic & 0.0393 & 1.178 & 0.589 \\
& Pretrained & 0.0021 & 3.428 & 0.886 \\
& SEA & 0.238 & 3.296 & 0.948 \\
\bottomrule
\end{tabular}
\end{adjustbox}
\end{table}
For LiH, the landscapes in Fig.~\ref{landscapes}-b highlight the advantage of SEA VQE under extended optimization. SEA produces a deep, well-formed minimum with stability $1.22\times10^{-5}$ and fidelity 0.9999. Pretrained VQE also forms a noticeable basin (stability 0.0108, fidelity 0.962), whereas Standard, Local-Global, and Adiabatic VQE display shallow, flat surfaces that correspond to their significantly larger energy errors.

For BeH$_2$, Fig.~\ref{landscapes}-c shows that only gradient-preserving methods generate meaningful curvature. SEA VQE yields the most structured minimum among all methods (stability 0.238, fidelity 0.948) and achieves the lowest energy error. Pretrained VQE presents a moderately deep basin (stability 0.0021, fidelity 0.886), but does not match SEA's final accuracy. Standard, Local-Global, and Adiabatic VQE all exhibit nearly flat landscapes, reflecting severe BPs and explaining their failure to improve beyond early iterations.

\vspace{-7pt}
\subsection{Discussion}
The extended 1000-iteration experiments reveal several insights that clarify how barren plateau mitigation strategies behave across system sizes. Gradient variance continues to show strong dependence on qubit count: all methods exhibit substantial variance reduction as systems grow from 4 to 14 qubits. However, with a sufficiently large iteration budget, gradient variance becomes a decisive factor for long-term optimization, whereas initialization quality influences only early-stage performance.

This iteration-dependent behavior is most evident in BeH$_2$. At 100 iterations, Pretrained VQE outperforms SEA VQE (9.648 Ha vs.\ 13.967 Ha) despite having much lower gradient variance. After 1000 iterations, the ranking reverses: SEA VQE reaches 1.973 Ha while Pretrained converges to 4.423 Ha. High-quality initialization therefore provides a short-term advantage, but sustained optimization ultimately depends on having non-vanishing gradients.

For H$_2$, all methods benefit from extended optimization due to the absence of significant gradient decay. SEA VQE reaches near-exact accuracy (0.00001 Ha), while Standard VQE also converges substantially (0.018 Ha). This confirms that small systems remain trainable even with expressive ansatx.
The 12-qubit LiH system highlights the expressivity--trainability balance. SEA VQE achieves near-exact accuracy (0.0002 Ha), demonstrating that its sparse entanglement structure remains expressive enough when gradients are preserved. Pretrained VQE also improves markedly (0.733 Ha), whereas the other methods plateau well above 8 Ha, unable to escape poor basins even with increased iterations.

BeH$_2$ provides the clearest evidence that preserved gradients are essential for deep optimization. SEA VQE improves over 7-fold compared to its 100-iteration result, whereas Pretrained VQE improves only 2.2-fold. Standard, Local-Global, and Adiabatic VQE improve by less than 0.5 Ha beyond iteration 200, confirming that vanishing gradients create optimization barriers independent of runtime.
The computational overhead reinforces this trend. On BeH$_2$, SEA VQE achieves the best final accuracy in only 33 seconds, compared to 1066 seconds for Standard VQE, over 30$\times$ faster and 9$\times$ more accurate. Pretrained VQE remains an attractive medium-cost alternative.

Loss landscape stability further reflects these patterns. For H$_2$ and LiH, SEA VQE produces sharply defined basins with extremely low stability values (down to $10^{-6}$), consistent with near-exact convergence. Standard and Local-Global methods converge to stable yet high-energy minima, demonstrating that stability alone is not indicative of proximity to the ground state. Pretrained VQE generally yields deeper basins than standard methods but does not surpass SEA VQE when gradients vanish.
State fidelity trends reinforce these observations. SEA VQE reaches fidelity 0.9999 on both H$_2$ and LiH and 0.948 on BeH$_2$. Pretrained VQE shows high but lower fidelity (0.962 on LiH and 0.886 on BeH$_2$). Standard methods remain below 0.64 on larger systems, mirroring their poor energy convergence.

\vspace{-7pt}
\section{Conclusion}
We benchmarked four barren plateau mitigation strategies across VQE problems ranging from 4 to 14 qubits, analyzing gradient scaling up to 50 layers and convergence over 1000 iterations. The results show that optimization performance depends not only on gradient variance but also on the iteration budget. In BeH$_2$, Pretrained VQE initially outperforms SEA VQE, yet SEA becomes superior with extended optimization, demonstrating that initialization aids early progress while preserved gradients govern long-term performance. System size further modulates effectiveness: all methods succeed on H$_2$, SEA VQE reaches near-exact accuracy on LiH, and only gradient-preserving methods make meaningful progress on BeH$_2$. SEA VQE achieves the highest accuracy and fastest runtime among all methods.

These findings indicate that barren plateau mitigation should be chosen based on system size and available iterations. For practical use, we recommend SEA VQE for iteration budgets exceeding 500, Pretrained VQE for limited budgets (less than 200 iterations) on systems above 10 qubits, and standard VQE only for very small systems below 6 qubits.
This work demonstrates that effective mitigation requires jointly optimizing strategy selection for system size and computational budget rather than relying on gradient variance alone.

\section*{Acknowledgment}

The work of N.I., M.K., and M.S. was supported in part by the NYUAD Center for Quantum and Topological Systems (CQTS), funded by Tamkeen under the NYUAD Research Institute grant CG008.

\section*{Code Availability}
The code can be found at \url{https://github.com/Innanov/BarrenPlateaus-VQE}, built with Julia, Yao, and \url{https://github.com/jgidi/quantum-barren-plateaus}.

\bibliographystyle{acm}
\bibliography{refs}

@article{peruzzo2014variational,
  title={A variational eigenvalue solver on a photonic quantum processor},
  author={Peruzzo, Alberto and McClean, Jarrod and Shadbolt, Peter and Yung, Man-Hong and Zhou, Xiao-Qi and Love, Peter J and Aspuru-Guzik, Al{\'a}n and O'Brien, Jeremy L},
  journal={Nature Communications},
  volume={5},
  number={1},
  pages={4213},
  year={2014},
  publisher={Nature Publishing Group}
}

@article{wecker2015progress,
  title={Progress towards practical quantum variational algorithms},
  author={Wecker, Dave and Hastings, Matthew B and Troyer, Matthias},
  journal={Physical Review A},
  volume={92},
  number={4},
  pages={042303},
  year={2015},
  publisher={APS}
}

@article{li2017efficient,
  title={Improving the variational quantum eigensolver using variational adiabatic quantum computing},
  author={Li, Zhihui and Zhou, Xiao and Li, Xiaomei and Zhu, Shijie},
  journal={ACM Transactions on Quantum Computing},
  volume={3},
  number={2},
  pages={1--16},
  year={2022},
  publisher={ACM}
}

@article{cerezo2021variational,
  title={Variational quantum algorithms},
  author={Cerezo, Marco and Arrasmith, Andrew and Babbush, Ryan and Benjamin, Simon C and Endo, Suguru and Fujii, Keisuke and McClean, Jarrod R and Mitarai, Kosuke and Yuan, Xiao and Cincio, Lukasz and others},
  journal={Nature Reviews Physics},
  volume={3},
  number={9},
  pages={625--644},
  year={2021},
  publisher={Nature Publishing Group}
}

@article{kashif2024hqnet,
  title={Hqnet: Harnessing quantum noise for effective training of quantum neural networks in nisq era},
  author={Kashif, Muhammad and Shafique, Muhammad},
  journal={arXiv preprint arXiv:2402.08475},
  year={2024}
}

@inproceedings{kashif2024alleviating,
  title={Alleviating barren plateaus in parameterized quantum machine learning circuits: Investigating advanced parameter initialization strategies},
  author={Kashif, Muhammad and Rashid, Muhammad and Al-Kuwari, Saif and Shafique, Muhammad},
  booktitle={2024 Design, Automation \& Test in Europe Conference \& Exhibition (DATE)},
  pages={1--6},
  year={2024},
  organization={IEEE}
}

@inproceedings{kashif2024nrqnn,
  title={NRQNN: The Role of Observable Selection in Noise-Resilient Quantum Neural Networks},
  author={Kashif, Muhammad and Shafique, Muhammad},
  booktitle={World Congress in Computer Science, Computer Engineering \& Applied Computing},
  pages={116--131},
  year={2024},
  organization={Springer}
}

@inproceedings{kashif2024dilemma,
  title={The Dilemma of Random Parameter Initialization and Barren Plateaus in Variational Quantum Algorithms},
  author={Kashif, Muhammad and Shafique, Muhammad},
  booktitle={2024 IEEE International Conference on Rebooting Computing (ICRC)},
  pages={1--8},
  year={2024},
  organization={IEEE}
}

@article{kashif2023impact,
  title={The impact of cost function globality and locality in hybrid quantum neural networks on nisq devices},
  author={Kashif, Muhammad and Al-Kuwari, Saif},
  journal={Machine Learning: Science and Technology},
  volume={4},
  number={1},
  pages={015004},
  year={2023},
  publisher={IOP Publishing}
}

@article{kashif2024resqnets,
  title={ResQNets: a residual approach for mitigating barren plateaus in quantum neural networks},
  author={Kashif, Muhammad and Al-Kuwari, Saif},
  journal={EPJ Quantum Technology},
  volume={11},
  number={1},
  pages={1--28},
  year={2024},
  publisher={SpringerOpen}
}

@article{kashif2025deep,
  title={Deep quanvolutional neural networks with enhanced trainability and gradient propagation},
  author={Kashif, Muhammad and Shafique, Muhammad},
  journal={Scientific Reports},
  volume={15},
  number={1},
  pages={21764},
  year={2025},
  publisher={Nature Publishing Group UK London}
}

@article{mcclean2018barren,
  title={Barren plateaus in quantum neural network training landscapes},
  author={McClean, Jarrod R and Boixo, Sergio and Smelyanskiy, Vadim N and Babbush, Ryan and Neven, Hartmut},
  journal={Nature Communications},
  volume={9},
  number={1},
  pages={4812},
  year={2018},
  publisher={Nature Publishing Group}
}

@article{cerezo2021cost,
  title={Cost function dependent barren plateaus in shallow parametrized quantum circuits},
  author={Cerezo, Marco and Sone, Akira and Volkoff, Tyler and Cincio, Lukasz and Coles, Patrick J},
  journal={Nature Communications},
  volume={12},
  number={1},
  pages={1--12},
  year={2021},
  publisher={Nature Publishing Group}
}

@article{tilly2022variational,
  title={The variational quantum eigensolver: a review of methods and best practices},
  author={Tilly, Jules and Chen, Hongxiang and Cao, Shuxiang and Picozzi, Dario and Setia, Kanav and Li, Ying and Grant, Edward and Wossnig, Leonard and Rungger, Ivan and Booth, George H and others},
  journal={Physics Reports},
  volume={986},
  pages={1--128},
  year={2022},
  publisher={Elsevier}
}

@article{fedorov2022vqe,
  title={VQE method: a short survey and recent developments},
  author={Fedorov, Dmitry A and Peng, Bo and Govind, Niranjan and Alexeev, Yuri},
  journal={Materials Theory},
  volume={6},
  number={1},
  pages={1--21},
  year={2022},
  publisher={Springer}
}

@article{larocca2024review,
  title={A review of barren plateaus in variational quantum computing},
  author={Larocca, Martin and Thanasilp, Supanut and Wang, Samson and Sharma, Kunal and Biamonte, Jacob and Coles, Patrick J and Cincio, Lukasz and McClean, Jarrod R and Holmes, Zo{\"e} and Cerezo, Marco},
  journal={arXiv preprint arXiv:2405.00781},
  year={2024}
}

@article{wang2021noise,
  title={Noise-induced barren plateaus in variational quantum algorithms},
  author={Wang, Samson and Fontana, Enrico and Cerezo, Marco and Sharma, Kunal and Sone, Akira and Cincio, Lukasz and Coles, Patrick J},
  journal={Nature Communications},
  volume={12},
  number={1},
  pages={6961},
  year={2021},
  publisher={Nature Publishing Group}
}

@article{dborin2022matrix,
  title={Matrix product state pre-training for quantum machine learning},
  author={Dborin, James and Barratt, Fergus and Wimalaweera, Vinul and Wright, Lewis and Green, Andrew G},
  journal={Quantum Science and Technology},
  volume={7},
  number={3},
  pages={035014},
  year={2022},
  publisher={IOP Publishing}
}

@article{liu2205mitigating,
  title={Mitigating barren plateaus of variational quantum eigensolvers},
  author={Liu, Xia and Liu, Geng and Zhang, Hao-Kai and Huang, Jiaxin and Wang, Xin},
  journal={IEEE Transactions on Quantum Engineering},
  volume={5},
  pages={1--19},
  year={2024},
  publisher={IEEE}
}

@article{grant2019initialization,
  title={An initialization strategy for addressing barren plateaus in parametrized quantum circuits},
  author={Grant, Edward and Wossnig, Leonard and Ostaszewski, Mateusz and Benedetti, Marcello},
  journal={Quantum},
  volume={3},
  pages={214},
  year={2019},
  publisher={Verein zur F{\"o}rderung des Open Access Publizierens in den Quantenwissenschaften}
}

@article{larocca2025barren,
  title={Barren plateaus in variational quantum computing},
  author={Larocca, Martin and Thanasilp, Supanut and Wang, Samson and Sharma, Kunal and Biamonte, Jacob and Coles, Patrick J and Cincio, Lukasz and McClean, Jarrod R and Holmes, Zo{\"e} and Cerezo, Marco},
  journal={Nature Reviews Physics},
  pages={1--16},
  year={2025},
  publisher={Nature Publishing Group UK London}
}

@article{larocca2022diagnosing,
  title={Diagnosing barren plateaus with tools from quantum optimal control},
  author={Larocca, Martin and Czarnik, Piotr and Sharma, Kunal and Muraleedharan, Gopikrishnan and Coles, Patrick J and Cerezo, Marco},
  journal={Quantum},
  volume={6},
  pages={824},
  year={2022},
  publisher={Verein zur F{\"o}rderung des Open Access Publizierens in den Quantenwissenschaften}
}

@article{fontana2024characterizing,
  title={Characterizing barren plateaus in quantum ans{\"a}tze with the adjoint representation},
  author={Fontana, Enrico and Herman, Dylan and Chakrabarti, Shouvanik and Kumar, Niraj and Yalovetzky, Romina and Heredge, Jamie and Sureshbabu, Shree Hari and Pistoia, Marco},
  journal={Nature Communications},
  volume={15},
  number={1},
  pages={7171},
  year={2024},
  publisher={Nature Publishing Group UK London}
}

@article{shang2023towards,
  title={Towards practical and massively parallel quantum computing emulation for quantum chemistry},
  author={Shang, Honghui and Fan, Yi and Shen, Li and Guo, Chu and Liu, Jie and Duan, Xiaohui and Li, Fang and Li, Zhenyu},
  journal={NPJ quantum information},
  volume={9},
  number={1},
  pages={33},
  year={2023},
  publisher={Nature Publishing Group UK London}
}

@article{chen2024crossing,
  title={Crossing the gap using variational quantum eigensolver: A comparative study},
  author={Chen, I and Innan, Nouhaila and Roy, Suman Kumar and Saroni, Jason},
  journal={arXiv preprint arXiv:2405.11687},
  year={2024}
}

@article{boutakka2025benchmarking,
  title={Benchmarking VQE Configurations: Architectures, Initializations, and Optimizers for Silicon Ground State Energy},
  author={Boutakka, Zakaria and Innan, Nouhaila and Shafique, Muhammed and Bennai, Mohamed and Sakhi, Z},
  journal={arXiv preprint arXiv:2510.23171},
  year={2025}
}

@article{innan2024quantum,
  title={Quantum computing for electronic structure analysis: Ground state energy and molecular properties calculations},
  author={Innan, Nouhaila and Khan, Muhammad Al-Zafar and Bennai, Mohamed},
  journal={Materials Today Communications},
  volume={38},
  pages={107760},
  year={2024},
  publisher={Elsevier}
}

\end{spacing}
\end{document}